\documentclass[conference,a4paper]{IEEEtran}
\IEEEoverridecommandlockouts
\usepackage{cite}
\usepackage{amsmath,amssymb,amsfonts}
\usepackage{algorithm}
\usepackage{algpseudocode}
\usepackage{graphicx}
\usepackage{textcomp}
\usepackage{blindtext}
\usepackage{subfigure}
\usepackage{float} 
\usepackage{subfigure}
\usepackage{xcolor}
\usepackage{diagbox}
\usepackage{slashbox}
\usepackage{wrapfig}
\usepackage{makecell}
\def\BibTeX{{\rm B\kern-.05em{\sc i\kern-.025em b}\kern-.08em
		T\kern-.1667em\lower.7ex\hbox{E}\kern-.125emX}}
\makeatletter

\begin{document}
	\title{Evaluating the Impact of Numerology and Retransmission on 5G NR V2X Communication at A System-Level Simulation}
	\IEEEpeerreviewmaketitle
	\author{\IEEEauthorblockN{Donglin Wang\IEEEauthorrefmark{2}, Pranav Balasaheb Mohite\IEEEauthorrefmark{3}, Qiuheng Zhou\IEEEauthorrefmark{1},  Anjie Qiu\IEEEauthorrefmark{2}, and Hans D. Schotten\IEEEauthorrefmark{2}\IEEEauthorrefmark{1}}
		\IEEEauthorblockA{\textit{\IEEEauthorrefmark{2}Rhineland-Palatinate Technical University of Kaiserslautern-Landau, Germany} \\
 		$\{$dwang,qiu,schotten$\}$@eit.uni-kl.de \\
             $\{$pmohite$\}$@rhrk.uni-kl.de \\} 
		\IEEEauthorblockA{\textit{\IEEEauthorrefmark{1}German Research Center for Artificial Intelligence (DFKI GmbH), Kaiserslautern, Germany} \\
		$\{$qiuheng.zhou,schotten$\}$@dfki.de}
	}
	\maketitle
% ====================================================================================
%structure of this paper
%mobility prediction part based on multiple algorithms
%handover times reduction by applying MP

%%%08.11.2022
\begin{abstract}
In recent years,  Vehicle-to-Everything (V2X) communication opens an ample amount of opportunities to increase the safety of drivers and passengers and improve traffic efficiency which enables direct communication between vehicles. The Third Generation Partnership Project (3GPP) has specified a 5G New Radio (NR) Sidelink (SL) PC5 interface for supporting Cellular V2X (C-V2X) communication in Release 16 in 2017.  5G NR V2X communication is expected to provide high reliability, extra-low latency, and a high data rate for vehicular networks.  In this paper, the newly introduced features of 5G NR standards like flexible numerology, variable Subcarrier Spacing (SCS), and allocated Physical Resource Blocks (PRBs) have been inspected in 5G NR V2X communications. Moreover, the 5G NR V2X data packet will be distributed to all nearby User Equipment (UE) by the Transmitter (Tx). However, there may be instances where certain UEs fail to receive the data packets in a single transmission. Unfortunately, the SL Tx lacks a feedback channel to verify if the Receivers (Rxs) have received the information. To meet the stringent reliability and latency requirements of C-V2X communication, we suggest and assess a retransmission scheme along with a scheme that incorporates varying resource allocations for retransmission in NR V2X communication. The effect of retransmission schemes on NR V2X communication systems has been detected. In the end, a system-level simulator complied with the 3GPP 5G NR SL standards, for NR V2X communication focusing on SL communication has been implemented, provides variable Medium Access Control Layer (MAC) configures, and the Packet Reception Ratio (PRR) has been used to represent the system-level performance for NR V2X communication.
Furthermore, this paper includes a set of preliminary system-level simulation results on SL communication using the developed simulator.
\end{abstract}

\begin{IEEEkeywords}
5G NR, C-V2X, Retransmission, System-Level Simulation
\end{IEEEkeywords}

\section{Introduction}
Vehicular networks have been studied for decades to provide communication services in vehicular scenarios equipped with wireless interfaces. The primary goal of vehicular networks was originally focused on enhancing road safety for users and optimizing traffic efficiency on roads. In 3GPP Release 14, the V2X specification was introduced, utilizing Long-Term Evolution (LTE) as the underlying technology to support C-V2X communication via SL \cite{3gpprel14}. Nevertheless, in recent times, the scope of vehicular networks has expanded beyond solely addressing road safety. Various use cases have emerged, aligning with the ongoing research and development efforts in the automotive industry, particularly in the realm of autonomous driving. Additionally, advancements in technology have facilitated the integration of infotainment applications within vehicular networks, necessitating high throughput and low latency communication capabilities for enhanced user experiences. Standardization efforts within the 3GPP are currently underway to establish a 5G NR framework for V2X communication. These activities involve defining innovative usage scenarios including Ultra-Reliable Low-Latency Communications (URLLC) and Enhanced Mobile Broadband (eMBB) services. The advanced features of 5G NR, coupled with its ability to cater to diverse use cases, position it as an optimal choice for C-V2X networks \cite{3gpprel16}\cite{3gpprel17}. 5G NR encompasses flexible numerologies and agile frame structure, in addition to a number of Resource Blocks (RBs), the number of Orthogonal Frequency Division Multiplexing (OFDM) symbols can also be variably allocated to a user, which in combination with wide-bandwidth operation, the usage of mmWave bands, and advanced multiple access techniques for C-V2X communication \cite{5gnrlsy2017}. 

\cite{5gnrv2xlinklevel2022} proposed a link-level simulation for 5G NR SL standards with some key parameters to evaluate the system performance.  In \cite{5gnrv2xcc2019}, the authors conducted an investigation and found that NR features offer several enhancements in terms of message reliability and timeliness, surpassing the performance of the legacy C-V2X solution. In this study, we not only analyze the system-level performance of 5G NR V2X communication with the incorporation of new NR features but also examine the impact of retransmission on the 5G NR V2X system.
\begin{figure}[htbp]
	\centering
	\includegraphics[width=\linewidth]{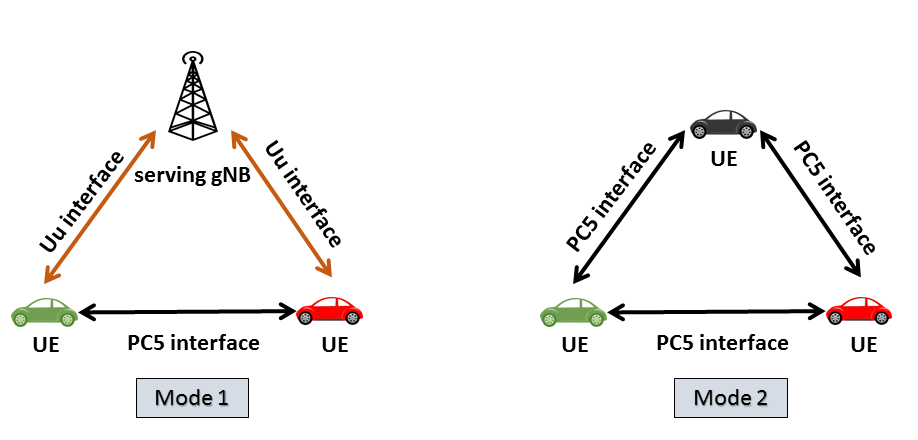}
	\caption{Transmission modes}
	\label{mode12}
\end{figure} 
In \cite{5gretranswang2020}, the authors introduced and assessed a retransmission scheme considering different traffic speeds, specifically designed for C-V2X communication. However, it is important to note that these schemes were based on LTE technology rather than 5G NR. \cite{its5gnrstand2020} \cite{5gv2xrel162020} introduced the concept of adaptive retransmissions in 5G NR V2X, which signifies that the retransmission mechanism in NR is designed to support not only blind retransmissions (i.e., retransmissions without associated HARQ feedback but also feedback-based retransmissions. This enables V2X services to cater to diverse Quality of Service (QoS) requirements. To the best of my knowledge, there has not been any implemented work on retransmission schemes specifically designed for 5G NR V2X communication.  This retransmission scheme is essential to NR V2X communication. In this study, we implement a HARQ blind retransmission scheme at the system-level of direct NR V2X communication with adaptive resource allocation schemes for fulfilling the stringent reliability requirements of 5G NR V2X communication, taking advantage of the NR features.

The paper is organized as follows: Section II presents an introduction to the new features of the 5G NR V2X communication. Following that, Section III provides the system-level simulation assumptions. Section IV demonstrates the simulation results and analysis. Lastly, Section V concludes the paper and outlines our future work plans.

\section{5G NR V2X new features}
The 5G NR V2X technology is anticipated and specifically designed to cater to diverse requirements across different use cases, surpassing the capabilities of LTE C-V2X. 5G NR introduces numerous novel features and functionalities that significantly enhance data rates, decrease latency, and improve Spectral Efficiency (SE) in V2X communication systems. The comparisons between LTE V2X and 5G NR V2X can be found in \cite{dlwiccc2023}.
Regarding 5G NR V2X, the envisioned modifications can be categorized into the following groups as shown in Tab. \ref{5gnr}:
\begin{table}[htbp]
\caption{Physical layer parameters of 5G NR V2X}
\begin{center}
	\begin{tabular}{|p{3.5cm}<{\centering}|p{4.5cm}<{\centering}|}
		\hline 
		Parameters&	          5G NR\\
		\hline
            Standards&            3GPP Rel-16/Rel-17 \\
		\hline
		Carrier frequency&    FR1: 410-7125 FR2: 24250-52600 (MHz)\\  
		\hline
		Supported channel bandwidths&    FR1:  5, 10, 15, 20, 25, 30, 40, 50, 60, 80, 90, 100  
        FR2: 50, 100, 200, 400 (MHz) \\  
		\hline
		SCS&               15, 30, 60, 120, 240  (KHz) \\  
		\hline
		Slot length&       14 symbols\\
		\hline
		Max number of subscarriers&      3300\\  
		\hline
            Modulation scheme  &  QPSK, 16QAM, 64QAM, 256QAM \\
            \hline
		Channel coding&    \thead{Data: LDPC coding \\ Control: Polar coding} \\
		\hline
		Latency&             1 ms \\
		\hline
		Communication types&     broadcast, groupcast, unicast \\
            \hline
	      SL	resource allocation &  Mode 1 and Mode 2    \\
		\hline
        CAM message &  periodic and aperiodic \\
           \hline
        SL physical channel & PSCCH, PSSCH, PSBCH, PSFCH \\
            \hline
        Retransmission & Blind and HARQ-feedback for unicast and groupcast retransmission \\
        \hline
	\end{tabular}
\end{center}
\label{5gnr}
\end{table}
\subsubsection{Frequency Range}
5G NR V2X SL can operate at wide frequencies, i.e., at frequencies within the two following frequency ranges \cite{fr1} \cite{fr2}: Frequency Range 1 (FR1) operates within the frequency range of 410 to 7125 MHz and offers channel bandwidth options of 5, 10, 15, 20, 25, 30, 40, 50, 60, 80, 90, and 100 MHz. Frequency Range 2 (FR2) on the other hand, spans from 24250 to 52600 MHz and provides channel bandwidth options of 50, 100, 200, and 400 MHz. Although both frequency ranges are supported in the 5G NR V2X SL, the design of the 5G NR V2X SL primarily focuses on FR1. No specific optimizations have been performed for FR2 in the NR V2X SL \cite{3gpprel16}. 

\subsubsection{Numerology}
In 5G NR, a flexible frame structure is employed to provide versatility. This flexibility is achieved through the use of different SCSs, which represents the distance between the centers of two consecutive subcarriers. The term "multiple numerologies" is used to describe this concept. The specifications introduce the parameter $\mu$, which denotes the numerology. The SCS is calculated using the formula: $ \Delta f = 2^\mu * 15 $, where $\mu$ is an integer that can be optimized for different scenarios. In 3GPP 5G NR Release 15 \cite{3gpprel15}, $\mu$ can take values from 0 to 4. Numerology serves as an indicator of the SCS. Therefore, 5G NR supports SCS values of 15 kHz, 30 kHz, 60 kHz, 120 kHz, and 240 kHz. In FR1, which corresponds to lower frequencies, only SCS values of 15 kHz, 30 kHz, or 60 kHz are used. The reason behind having multiple SCS options in NR is to enable the deployment of 5G in both lower and higher frequency bands \cite{dlwiccc2023}.

% \begin{table*}[htbp]
% \caption{Multiple Numerologies in NR \cite{dlwiccc2023}}
% \begin{center}
% 	\begin{tabular}{|p{1.5cm}<{\centering}|p{1.5cm}<{\centering}|p{2cm}<{\centering}|p{1.5cm}<{\centering}|p{1.5cm}<{\centering}|p{2cm}<{\centering}|p{2cm}<{\centering}|}
% 		\hline 
% 		CP&	        SCS [KHz]& $\#$subframes per radio frame&   $\#$slots per subframes&  slot duration(ms)& $\#$OFDM symbols per slot& 	 Applicable frequency range \\
% 		\hline
%             \hline
% 		normal&		15&	       10&	 1& 	1& 14& FR1\\  
% 		\hline
% 		normal&		30&     	10&	 2&	    0.5& 14& FR1\\  
% 		\hline
% 		normal&		60&	        10&	 4&	    0.25& 14& FR1 and FR2\\  
% 		\hline
% 		normal&		120&        10&  8&      0.125& 14& FR2\\  
% 		\hline
% 	\end{tabular}
% \end{center}
% \label{mu}
% \end{table*}

\subsubsection{Transmission Modes}
In this section, the transmission modes of 5G NR SL communications are
explained. Similar to LTE V2X Mode 3 and Mode 4 \cite{wang2019methodologies}, there are 2 modes of
operation namely Mode 1 and Mode 2 for 5G NR V2X communications.
\begin{itemize}
    \item SL Transmission Mode 1
\end{itemize}
Just like LTE Mode 3, SL Transmission Mode 1 is applicable when vehicles are within cellular coverage. In this mode, the resource allocation for communication is facilitated through the cellular infrastructure, such as the gNB (5G base station). To aid the resource allocation process at the gNBs, the UE context information, including traffic pattern geometrical information, can be reported to the gNBs. As depicted in Fig. \ref{mode12}, UEs exchange data packets directly with each other via the PC5 interface. These UEs remain within the communication range of the serving gNB which assigns the SL radio resources for V2X communications via through the Uu interface.
\begin{itemize}
    \item SL Transmission Mode 2
\end{itemize}
5g NR V2X Mode 2 is a communication mode where direct data transfer between vehicles occurs through the PC5 air interface. This mode does not rely on the presence of gNB for resource scheduling. It enables communication between vehicles even when they are not within cellular coverage. UEs have the capability to autonomously select their SL resources from a pool of available resources. These resources can consist of one or multiple sub-channels. This autonomous resource selection enables UEs to efficiently utilize the available resources based on their specific communication requirements and network conditions. NR V2X Mode 2 operates similarly to LTE V2X Mode 4, which utilizes the Semi-Persistent Scheduling (SPS) scheme for resource scheduling \cite{garcia2021tutorial}.

The direct 5G NR V2X communication utilizing transmission Mode 1 is employed. This indicates that all UEs involved in the simulation are within the coverage of a cellular network.

\subsubsection{Communication Types and New Physical SL Feedback Channel}
In order to cater to a wide range of applications, NR V2X supports three types of transmissions: broadcast, groupcast, and unicast \cite{3gpprel16}. These transmission modes allow for efficient communication between vehicles in various scenarios.
In 5G NR V2X communication, there are two additional channels called the Physical SL Shared Channel (PSSCH) and the Physical SL Control Channel (PSCCH). These channels serve specific purposes in SL communication. A UE can be configured by higher layers with one or more SL resource pools. A SL resource pool is a dedicated set of resources allocated for the transmission of PSSCH. The PSSCH is used for transmitting data and information between UEs engaged in SL communication. Each transmission on the PSSCH is associated with a corresponding transmission on the PSCCH. The PSCCH is responsible for carrying control information related to the PSSCH transmissions. 
Additionally, 5G NR V2X introduces a new channel called the Physical Sidelink Feedback Channel (PSFCH), which is a distinction from LTE C-V2X where it is not considered. The PSFCH is utilized for transmitting HARQ feedback and is specifically required for unicast and groupcast transmission modes. A UE can only perform HARQ retransmissions for groupcast or unicast transmissions if a PSFCH is (pre-)configured in the resource pool \cite{3gpprel1638885}. 

There are more additional features of 5G NR V2X communications that can be found in Tab. \ref{5gnr}. 
\begin{table}[htbp]
\caption{Simulation parameters}
\begin{center}
\begin{tabular}{ |p{3.5cm}||p{3.5cm}| }
 \hline
 \textbf{Parameters}  &  \textbf{Value} \\
 \hline
 \hline
 Highway length & 5196 meters \\
 \hline
 Highway width & 4 meters \\
 \hline
 Inter-Site Distance (ISD) & 1732 meters \\
 \hline
 Deployment & 3 lanes in each direction \\
 \hline
 Inter-Vehicle Distance (IVD) & 3, 5, 10, 20, 40, 80, 100 meters \\
 \hline
 Carrier frequency & 5.9 GHz\\
 \hline
 Bandwidths& 10, 20 MHz\\
 \hline
 Transmission frequency& 10, 20, 30, 40, 50 Hz\\
 \hline
 Number of gNB& 3 \\
 \hline
 Height of gNB antenna  & 35 meters \\
 \hline
 Height of UE antenna & 1.5 meters \\
 \hline
 Communication range & 500 meters \\
 \hline
 Packet size& 300 bytes\\
 \hline
 Tx antenna Gain & 0 dB \\
 \hline
 Rx antenna Gain & 3 dB \\
 \hline
 UE transmission power& 24 dBm\\
\hline
 SCS & $\mu$=0 (15 kHz), $\mu$=1 (30 kHz), $\mu$=2 (60 kHz)\\
 \hline
  Thermal noise & -174 dBm\\
 \hline
  Channel mode & WINNER II \\
 \hline 
 Fading + shadowing & Enabled\\
 \hline
 Load of interfering gNBs & 100\% (Full buffer)\\
 \hline 
 Transmission frequency & 10, 20, 30 Hz\\
 \hline
\end{tabular}
\end{center}
\end{table}

\section{system-level Simulation Assumptions}

\begin{figure}[htbp]
	\centering
	\includegraphics[width=\linewidth]{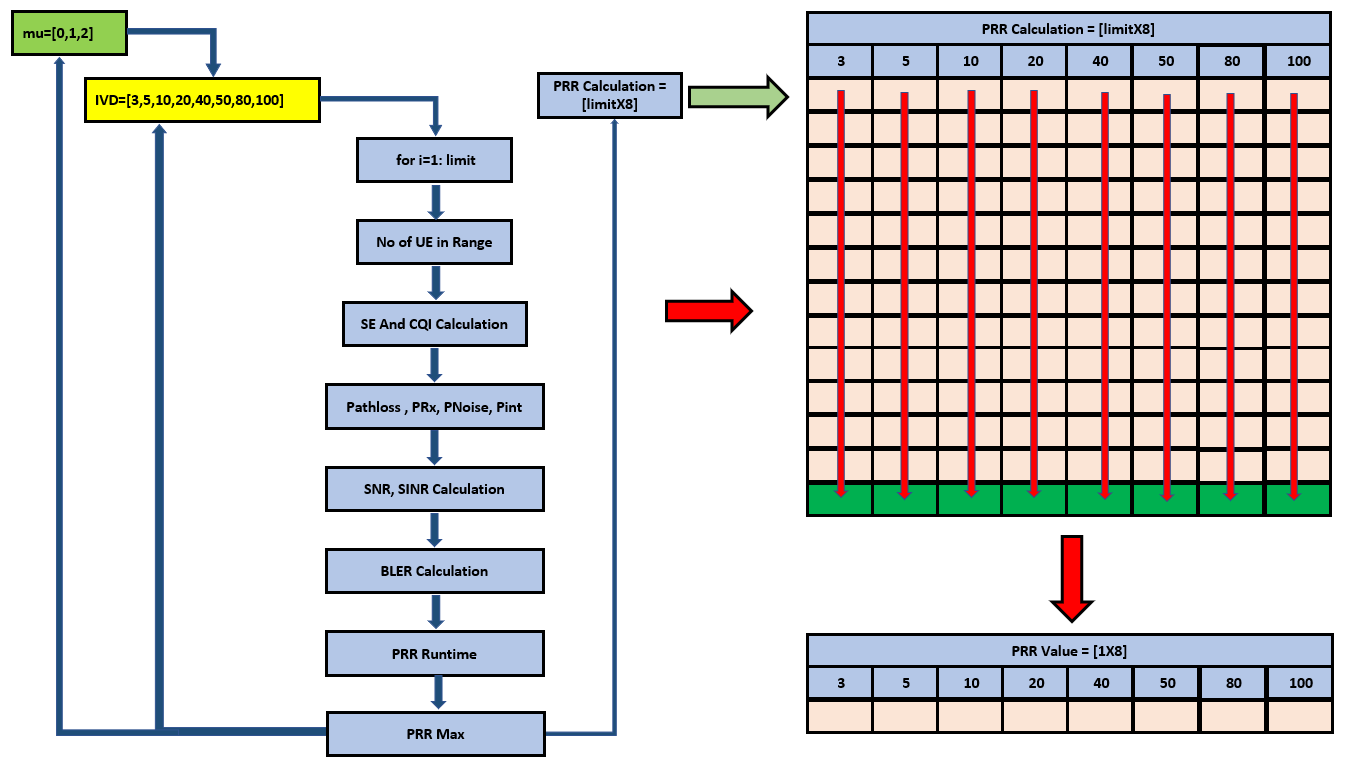}
	\caption{Methodology of system-level simulation}
	\label{slflowchart}
\end{figure} 
The 5G NR V2X system-level simulator is implemented in MATLAB that generates specific outputs based on given input conditions. It takes into account factors such as the number of vehicles under the coverage of the gNB, pathloss values, interference power, SINR, Modulation and Coding Scheme (MCS), and Block Error Rate (BLER) calculations. To evaluate the system's performance, a Key Performance Indicator (KPI) known as the PRR is considered. The methodology and simulation steps of the system-level simulation without retransmission are outlined and depicted in Fig. \ref{slflowchart}.

% PRR provides insights into the quality and reliability of the communication links within the system. The analysis of performance primarily involves examining the variation of PRR values for different Inter-Vehicle Distances (IVDs) and numerology values, both with and without blind retransmission. 

\subsection{Simulation Scenario}
In our simulation environment, the highway consists of six lanes, with three lanes in one direction. The total length of the highway is 5196 meters. For network infrastructure, three gNBs are deployed along the highway at an Inter-Site Distance (ISD) of 1732 meters, as specified in \cite{3gpprel1537885}.

\subsection{Number of UEs}
The UEs on the highway scenario all travel at an identical speed and have the same Inter-Vehicle Distance (IVD) value. The total number of UEs ($UE_H$) present in the highway scenario has been computed as follows:

\begin{equation}
    UE_H=\frac{L_H}{IVD} \times Lane \label{uehighway}
\end{equation}
where $L_H$ represents the length of the highway scenario, $Lane$ denotes the number of highway lanes, and $IVD$ defines it as before.
The calculation for determining the number of UEs ($UE_{gNB}$) controlled by each gNB is as follows:

\begin{equation}
    UE_{gNB}=\frac{ISD}{IVD} \times Lane \label{uehighway}
\end{equation}
where $ISD$, $IVD$ and $Lane$ are defined before. 

\subsection{Modulation and Coding Scheme}

SE calculation: SE is a metric that quantifies the utilization of the frequency spectrum. It represents the maximum amount of data that can be transmitted to a specified number of users per second while maintaining an acceptable quality of service. The formula used to calculate $SE$ is as follows:

\begin{equation}
    SE=\frac{P_s \times UE_{gNB} \times f_t}{BW} \label{se}
\end{equation}
Where $P_s$ is the packet size, $UE_{gNB}$ is the number of UE in range and $f_t$ is the transmission frequency. $BW$ is the transmission bandwidth. 

MCS selection: MCS selection in 5G NR is a procedure that varies depending on the implementation. 5G NR includes the Channel Quality Indicator (CQI), which is reported by the UE and can be utilized by the gNB for selecting the appropriate MCS index. NR defines three tables of 4-bit CQIs (see Tables 5.2.2.1-1 to 5.2.2.1-3 in \cite{3gpprel1638214}), and each table is associated with one MCS table. Table 5.2.2.1-2: 4-bit CQI Table is utilized in this work. 

\subsection{Channel Model}
The channel model is a mathematical representation of the effects experienced by the signal while propagating through the medium. Pathloss is the reduction or attenuation of electromagnetic signal carrying data from Tx to Rx. We have used the Winner II channel model for pathloss calculation \cite{3gpprel14}.

\subsection{Interference}
In SL 5G NR V2X communication, the data packets are directly transmitted from a Tx to an Rx without traversing the network infrastructure. The gNBs in the cellular network are primarily responsible for providing control information to the UE within their coverage area. The interference in this scenario arises from other Txs that are also utilizing the same transmission bandwidth resource and transmitting data packets concurrently.
\begin{figure}[htbp]
	\centering
	\subfigure[delta=0dB]{
		\includegraphics[width=0.1\textwidth]{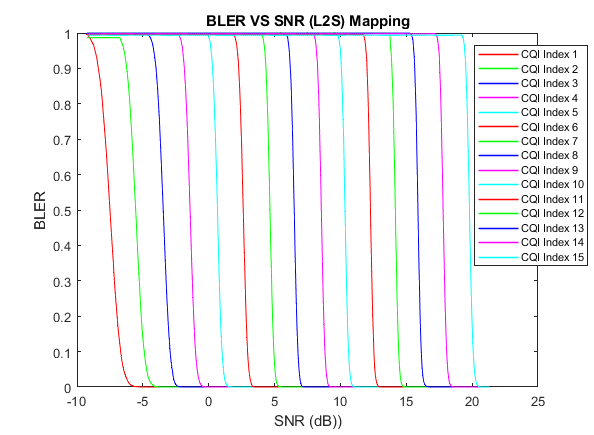}}
	\label{delta0}
         \subfigure[delta=3dB]{
		\includegraphics[width=0.1\textwidth]{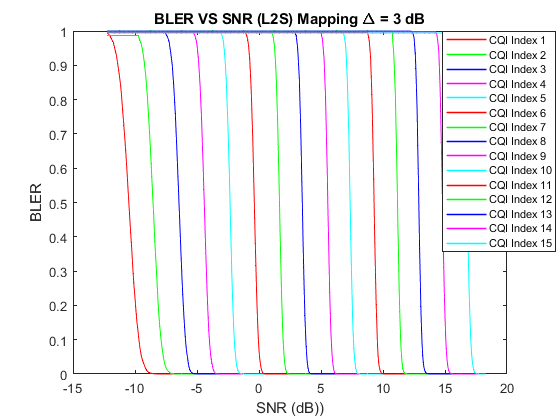}}
	\label{delta5}
        \subfigure[delta=5dB]{
		\includegraphics[width=0.1\textwidth]{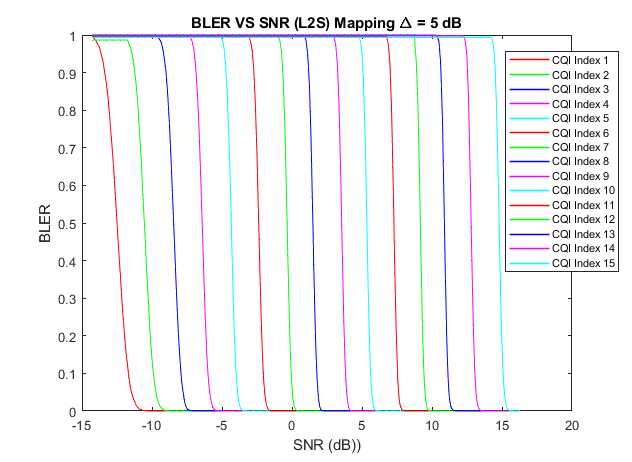}}
	\label{delta5}
        \subfigure[delta=7dB]{
		\includegraphics[width=0.1\textwidth]{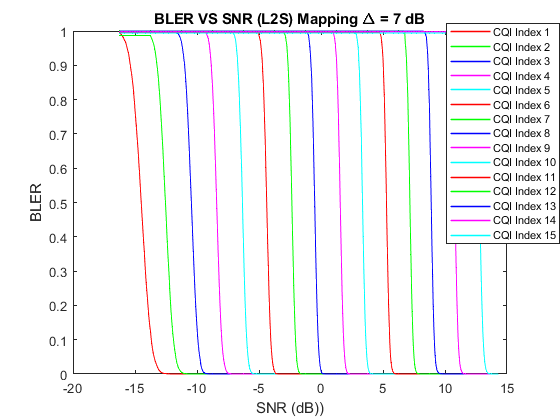}}
	\label{delta7}
    \caption{BLER and SNR curves}
    \label{resultsforscenarios}
\end{figure}
\subsection{BLER and SINR Mapping}
During system-level simulations, when dealing with a significant number of nodes, the detailed modeling of PHY processing can be substituted with an abstraction model implemented through a Link-to-System Mapping (L2SM) interface. In this approach, the complexity of the PHY layer is simplified, allowing for more efficient and scalable simulations. As shown in Fig. \ref{resultsforscenarios}(a), the mapping table considers all MCS (1-15) for the ITU-Extended Vehicular A (EVA) channel fading, which is denoted as L2SM-0 \cite{viennaltesimu}. The SINR value $x_i$ is calculated for each Rx within the communication range for a specific Tx. The corresponding BLER is obtained by referencing the L2S-0 mapping, and if the SINR value falls outside the range covered by the link simulation, a constant-value extrapolation technique is used. To determine if a packet is received successfully, a random number (X) is generated from a uniform distribution between 0 and 1. If X is less than the BLER value, the packet is marked as received. 

In our simulations, we considered different versions of BLER curves for later retransmission, which were adjusted by a parameter $\Delta$  to model the improved sensitivity of a 5G receiver. In \cite{nardini2020simu5g}, Figure 14 displays the BLER curves obtained with $\Delta$ values of 3 dB, 5 dB, and 7 dB, denoted as L2SM-3, L2SM-5, and L2SM-7 respectively for retransmission in our work.

\subsection{PRR Calculation}

In the considered scenario, the traffic consists of Cooperative Awareness Messages (CAM), where each message is encapsulated in a single packet, which, in turn, is contained within a single transport block. As a result, the PRR can be obtained simply as the inverse of the BLER. The PRR is calculated by considering all the Rxs within the intended communication range of the Tx. For each transmitted CAM message, the PRR can be computed as $N/M$, where $M$ represents the total number of UEs located within the communication range of the UE transmitting the message, and $N$ denotes the count of UEs within that range that successfully receive the message. 

\subsection{PRR Calculation for Overloaded Scenarios}

In our simulation, all gNBs are assumed 100\% loaded. In overloaded scenarios where the gNB is unable to support all UEs, some UEs may be dropped due to resource limitations. By incorporating the dropped UEs in the PRR calculation, a more comprehensive assessment of the system performance in overloaded conditions can be obtained, allowing for a better understanding of the impact on packet reception and overall system reliability.

In such cases, the PRR calculation takes into account the dropped UEs. The maximal achievable PRR in an overloaded scenario, denoted as $PRR_{max}$, is calculated as the ratio of the total number of supported UEs ($UE_{supported}$) to the total number of UEs within the coverage area of the gNB as follows:
\begin{equation}
    PRR_{max}= \frac{UE_{supported}}{UE_{gNB}} \label{prrmax}
\end{equation}
where $UE_{supported}$ is essential element to calculate. If the system is not overloaded, the $PRR_{max}$ is always 1. 
\begin{table}[htbp]
\caption{PRBs for given BW of different numerology}
\begin{center}
\begin{tabular}{ |p{2.5cm}|p{0.5cm}|p{0.5cm}|p{0.5cm}|p{0.5cm}|  }
\hline
\backslashbox[29mm]{\textbf{SCS}}{\textbf{PRBs}} {\textbf{BW (MHz)}}
&\makebox{\textbf{5}}&\makebox{\textbf{10}}&\makebox{\textbf{15}}&\makebox{\textbf{20}} \\
\hline
15 kHz (mu=0)   & 25 & 52  & 79 & 106 \\ \hline
30 kHz (mu=1)   & 11 & 24  & 38 & 51  \\ \hline
60 kHz (mu=2)   & NA & 11  & 18 & 24  \\ \hline
\end{tabular}
\label{PRBs}
\end{center}
\end{table}

As shown in Tab. \ref{PRBs}, the number of PRBs for a particular BW and numerology are calculated. For example, if the BW is 10 MHz, $mu=0$ and $mu=1$ are used,  we can find the SCS value has increased from 15 kHz to 30 kHz, and the PRBs are decreased from 52 to 24 respectively. Then we consider the PRBs allocated for mentioned PSCCH and PSSCH. The PRB allocated for PSCCH ($NPRB_{PSCCH}$) is fixed to 2. On the other hand, the PRBs allocated for PSSCH are calculated as: 
\begin{equation}
NPRB_{PSSCH} = ceil(\frac{P \times 8}{N_{sb}\times N_{us}\times Max_{mcs}})    \label{nprb}
\end{equation}
where $P$ is the packet size, $N_{sb}$ is the number of subcarriers in one resource block which is related to numerology value $u$, $N_{us}$ represents the number of usable symbols which is 9 in this simulation, and $Max_{mcs}$ denotes the maximum data rate efficiency. 

The total number of PRBs needed to transmit a single message is the sum of the PRBs required for the transmission of both PSSCH and PSCCH:
\begin{equation}
NPRB_{total} =  NPRB_{PSSCH} + NPRB_{PSCCH} \label{nprbsum}
\end{equation}
$NPRB$ is an variable in Tab. \ref{PRBs}. 

The number of UEs ($UE_{sf}$) that can be scheduled in one subframe is then given as follows:
\begin{equation}
UE_{sf} =  \frac{N_{PRB}}{NPRB_{total}} \label{UEspportedsf}
\end{equation}

Therefore, the number of UE supported ($UE_{supported}$) per second is calculated as:
\begin{equation}
UE_{supported} =  floor (\frac{UE_{sf} \times N_{sf}}{T})\label{uesupported}
\end{equation}
where $N_{sf}$ is the number of subframes, and the $T$ is the transmission frequency. 

The PRR is calculated during simulation runtime, taking into account only the $UE_{supported}$. We denote the runtime PRR ($PRR_{runtime}$) which is a percentage of UEs that successfully receive a packet from the tagged Tx among the Rxs within the transmission range of the Tx in the running time as we defined before. the final effective PRR is calculated as: 
\begin{equation}
PRR =  PRR_{max}* PRR_{PRR_{runtime}}\label{PRR}
\end{equation}
\subsection{Resource Allocation for Retransmission}

\begin{figure}[htbp]
	\centering
	\includegraphics[width=\linewidth]{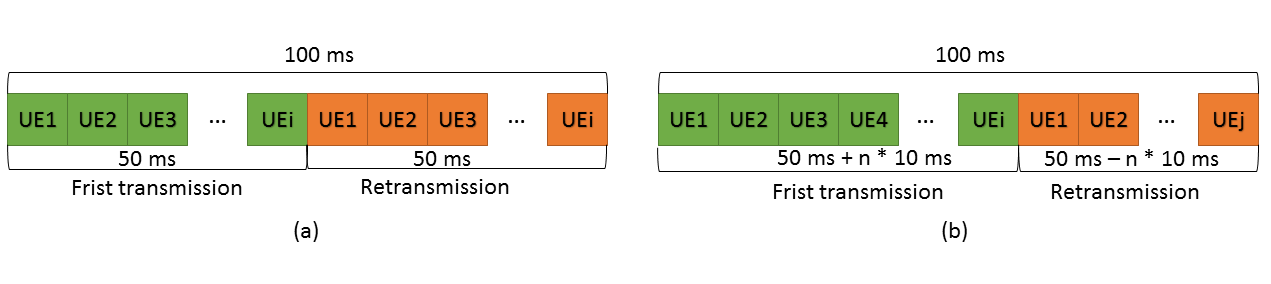}
	\caption{Resource allocation schemes \cite{5gretranswang2020}}
	\label{resourceallocation}
\end{figure}
In this simulation, to enhance the performance of direct NR V2X communication, single-blind retransmission is introduced. HARQ feedback is essential for unicast and groupcast transmission modes, but for the purposes of this study, we focus solely on broadcast transmission.
\begin{itemize}
    \item Equal resource allocation scheme for retransmission
\end{itemize}
In the illustrated example depicted in Fig. \ref{resourceallocation}(a),  $U_1, U_2, \cdots, U_i$ are being served by a gNB within a time duration of 100 ms, corresponding to one period of 10 Hz. The resource allocation scheme assumes that both the initial transmission and retransmission each require 50 ms of resources. The number of UEs, represented by the variable $i$, is supported by the gNB based on factors such as IVDs and message rate. 

As mentioned before $x_1$ is the calculated SINR value for the first transmission. Then a new SINR value $x_2$ of retransmission is calculated for each Rx in the communication range for the same selected Tx. Both of these SINR values are averaged together to obtain the final SINR value ($x_{final}$) as follows:
\begin{equation}
    x_{final} = mean(x_1,x_2) \label{sinrvalue}
\end{equation}
The corresponding BLER is obtained by referring to the L2SM-3, L2SM-5, and L2SM-7 tables. This lookup process is similar to the previous steps described.

\begin{itemize}
    \item Nonequal resource allocation scheme for retransmission
\end{itemize}

The resource allocation process is based on the assumption that the allocation of packet transmission resources is not evenly divided between the initial transmission and retransmission. Fig. 4(b), $U_1, U_2,  \cdots, U_i$ are being served by a gNB within a time duration of $50+n*10$ ms. $U_1, U_2,\cdots, U_j$ are being served by a gNB within a time duration of $50-n*10$ ms.
As a result, the transmission and retransmission are performed using different CQIs index values for a given interference, noise, and IVD condition.
Consequently, the PRR calculation is carried out separately for the initial transmission and the retransmission stages. These individual PRR values are then averaged together to obtain the final PRR. Mathematically, the PRR calculation can be expressed as follows:

\begin{equation}
    PRR = mean(PRR_1, PRR_2) \label{prrvalue}
\end{equation}

\begin{figure*}[htbp]
	\centering
	\subfigure[Numerologies vs. IVD]{
		\includegraphics[width=0.33\textwidth]{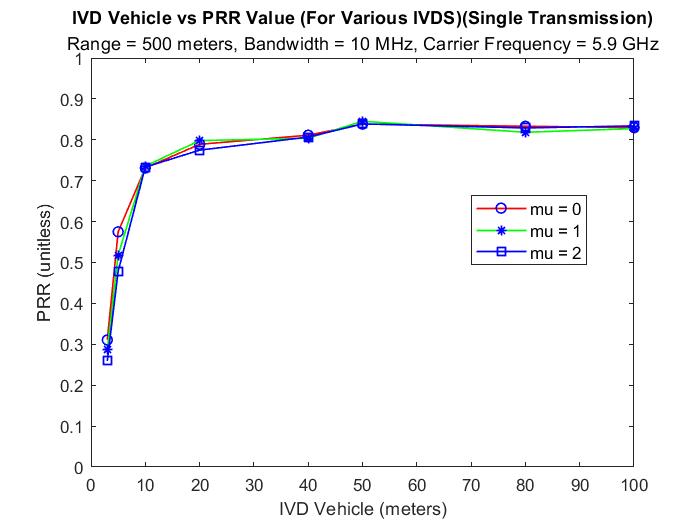}}
	\label{muvsivd}
         \subfigure[TF vs. IVD]{
		\includegraphics[width=0.34\textwidth]{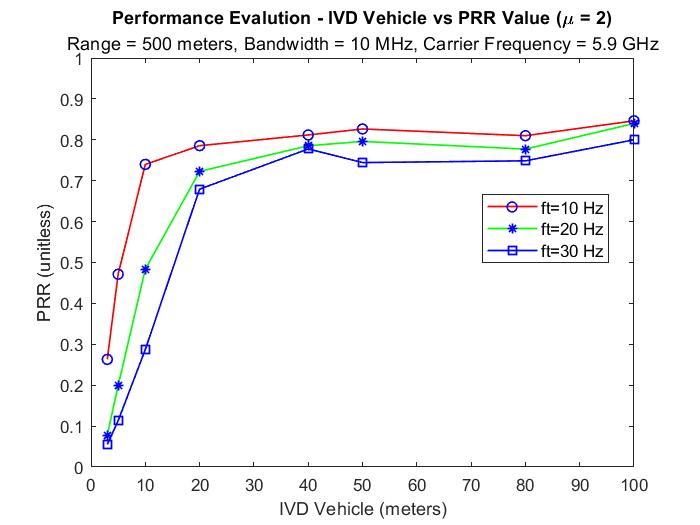}}
	\label{mu2vstf}
        \subfigure[Retransmission with different mapping tables]{
		\includegraphics[width=0.33\textwidth]{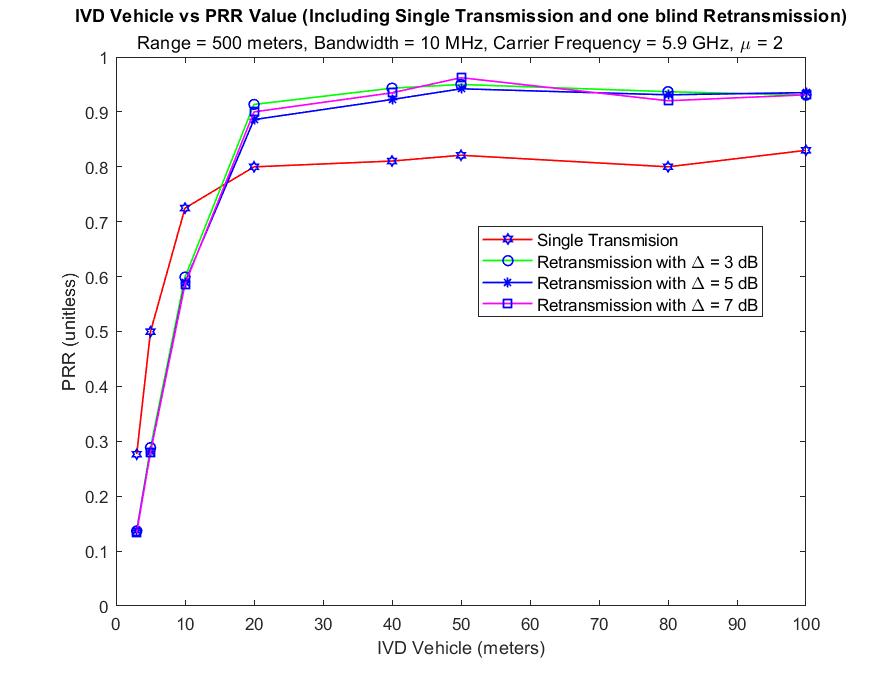}}
	\label{mu2retrans}
        \subfigure[Adaptive resource allocation scheme]{
		\includegraphics[width=0.36\textwidth]{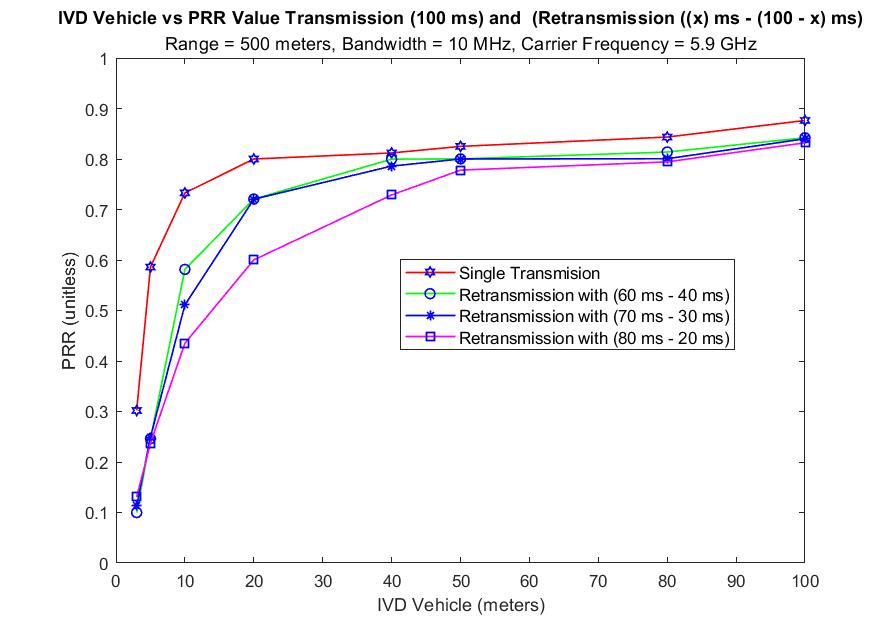}}
	\label{mu0retrans}
    \caption{PRR vs. IVD with and without retransmission}
    \label{results}
\end{figure*}

\section{Results Analysis}
We conducted a simulation of the direct NR V2X communication system, considering different IVD values, numerologies, and Transmission Frequencies (TFs). Additionally, we evaluated the performance of the NR V2X communication system with retransmission using two different resource allocation schemes. The results of these simulations are presented and analyzed in this study. 

\subsubsection{Evaluation of various numberologies}
As shown in Fig. \ref{results}(a), to examine the impact of numerology ($\mu$) on the performance of direct NR V2X communication, we analyzed PRR values for different $\mu$ values (0, 1, and 2) across various IVDs. For instance, when considering a numerology $\mu$ value of 0, we observed that the PRR increased from 31.06\% to 83.00\% as the IVD values ranged from 10 to 100 meters. Similarly, for a $\mu$ value of 1, the PRR increased from 28.62\% to 82.74\%. These findings indicate that as the IVD increases, fewer vehicles are present in the highway scenario, resulting in reduced system load and an MCS with better robustness applied. Consequently, the NR V2X communication system performance improves with lower IVD values. The subsequent step involves examining the PRR curves corresponding to the numerologies that were tested. This observation can be attributed to the inherent flexibility introduced by each numerology in terms of the number of subchannels available in the frequency domain and the duration of slots in the time domain. In the NR system, the processing times and transmission durations exhibit an inverse relationship with the SCS. Specifically, a lower SCS results in a higher number of PRBs within a given bandwidth. Conversely, a higher SCS leads to shorter slot durations, implying reduced timings. Our analysis revealed that the PRR curves for the three numerologies closely overlap. This can be attributed to the fact that as SCS increases from $\mu=0$ to $\mu=1$, the number of available subchannels decreases. However, at the same time, the transmission time doubles, compensating for the reduced subchannels. As a result, the same amount of data packets can still be transmitted, leading to similar system performance across different numerologies. However, when $IVD <= 10$ meters, our simulation indicates that the system becomes overloaded. In this scenario, we observed a decrease in the PRR as the numerology ($\mu$) values increase. For instance, with a fixed IVD of 10 meters and a bandwidth of 10 MHz, the PRR value is 58\% for $\mu=0$, but it drops to 52.2\% for $\mu=1$. This behavior can be attributed to the fact that the available PRBs, ranging from 52 to 24, do not undergo a doubling reduction as the SCS values increase from 15 kHz to 60 kHz as shown in Tab. \ref{PRBs}. Meanwhile, the transmission time doubles. Consequently, the reduction in the number of available PRBs leads to a decrease in the number of UEs that can be supported. Thus, the system's performance is negatively impacted when the SCS value increases in an overloaded system.
\subsubsection{Evaluation of various transmission frequency}
To examine the impact of TF on the PRR, we investigated TF values of 10 Hz, 20 Hz, and 30 Hz while keeping the numerology ($\mu$) fixed at 2. Fig. \ref{results}(b) displays the results. When the IVD value is 20 meters, we observed PRR values of 79.1\% for a TF of 10 Hz, 73\% for a TF of 20 Hz, and 67.5\% for a TF of 30 Hz. According to Equation \ref{se}, a larger TF results in a higher required SE for the transmission technology. Consequently, in this case, an MCS with lower robustness may be employed, leading to a decrease in the performance of NR V2X communication.
\subsubsection{Evaluation of equal and nonequal resource allocation schemes for retransmission}
For equal resource allocation, as shown in Fig. \ref{results}(c), for $IVD >= 20$ meters, with a single retransmission, the system load naturally doubles, leading to the utilization of a less robust MCS. However, due to the application of different L2SM tables with offsets ($\Delta$), our simulation scenarios with retransmissions were able to achieve PRR above the target of 90\% when the system capacity was not overloaded. However, in cases where the system is overloaded, we observed that the single transmission outperforms the retransmission case. For instance, with an IVD of 10 meters, the PRR decreased from 74.2\% without retransmission to 60\% with one blind retransmission using L2SM-3. In this simulation, we also introduced nonequal resource allocation for the one blind retransmission in NR V2X communication. Fig. \ref{results}(d) illustrates the results, we found that the equal resource allocation for both transmission and retransmission yielded better performance compared to transmission and retransmission with nonequal resource allocation. This can be attributed to the fact that in the considered scenario, the interference power is significantly lower than the noise power. Consequently, allocating equal resources for both transmission instances proves to be the most effective strategy for maximizing performance.
\section{conclusion}
In this study, we have developed a system-level simulator to investigate the performance of direct 5G NR V2X communication in a highway scenario. The simulator considers various numerologies, TFs, and resource allocation schemes. The simulation results demonstrate the effect of flexible numerologies on the performance of direct NR V2X communications. It is observed that flexible numerology enables UEs to effectively utilize frequency domain diversity. However, the impact of different numerologies on the overall system performance of the direct NR V2X communication is found to be minimal. Additionally, to enhance the efficiency of retransmission in direct NR V2X communication, we have conducted a comprehensive analysis of different resource allocation schemes. The findings suggest that, in certain scenarios, retransmission can significantly improve system performance and achieve a PRR above 90\%. Future simulations will focus on further exploring NR V2X communication to gain deeper insights into its capabilities and potential improvements.
\section{acknowledgement}
This work has been supported by the Federal Ministry of Education and Research of the Federal Republic of Germany (BMBF) as part of the Open6GHub project with funding number 16KISK004. The authors would like to appreciate the contributions of all AI4Mobile partners. The authors alone are responsible for the content of the paper which does not necessarily represent the project. 

\bibliographystyle{IEEEtran}
\bibliography{references}

\end{document}